\documentclass[final,11pt,3p]{elsarticle}

\usepackage{graphicx} 
\usepackage{amsmath} 
\usepackage{lineno,hyperref}
\usepackage{setspace}
\begin{document}
\begin{frontmatter}
\title{Pulse-shape discrimination and energy quenching of alpha particles in Cs$_2$LiLaBr$_6$:Ce$^{3+}$}
\author[lanl]{K.E.~Mesick\corref{cor}}
\ead{kmesick@lanl.gov}
\author[lanl]{D.D.S.~Coupland}
\author[lanl]{L.C.~Stonehill}
\cortext[cor]{Corresponding author}
\address[lanl]{Los Alamos National Laboratory, Los Alamos, NM 87545 USA}

\begin{abstract}
Cs$_2$LiLaBr$_6$:Ce$^{3+}$ (CLLB) is an elpasolite scintillator that offers excellent linearity and gamma-ray energy resolution and sensitivity to thermal neutrons with the ability to perform pulse-shape discrimination (PSD) to distinguish gammas and neutrons.  Our investigation of CLLB has indicated the presence of intrinsic radioactive alpha background that we have determined to be from actinium contamination of the lanthanum component.  We measured the pulse shapes for gamma, thermal neutron, and alpha events and determined that PSD can be performed to separate the alpha background with a moderate figure of merit of 0.98.  We also measured the electron-equivalent-energy of the alpha particles in CLLB and simulated the intrinsic alpha background from $^{227}$Ac to determine the quenching factor of the alphas.  A linear quenching relationship $L_{\alpha} = E_{\alpha} \times q + L_0$ was found at alpha particle energies above 5~MeV, with a quenching factor $q = 0.71$~MeVee/MeV and an offset $L_0 = - 1.19$~MeVee.
\end{abstract}


\end{frontmatter}

\section{Introduction}

Elpasolite scintillators are a promising new class of inorganic crystals that have the ability to detect and distinguish between gamma-ray and neutron radiation.  This makes them an attractive option for applications that require excellent performance under constraints such as size and weight, for example handheld radiation detectors that support homeland security or space-based applications such as planetary science.  The elpasolites exhibit excellent energy linearity and high light output resulting in good energy resolution, and Li-containing elpasolites have good thermal neutron detection efficiency through the $^6$Li(n,$\alpha$)$^{3}$H capture reaction.  Cerium-doped Cs$_2$LiLaBr$_6$ (CLLB) is a more recently studied elpasolite with a density of $\rho = 4.2$~g/cm$^3$ that emits at a peak wavelength of 420~nm.  CLLB offers the highest light yield (up to 60,000 photons/MeV) and the best energy resolution at 662~keV (2.9\%) among the reported elpasolites~\cite{glodo2011_bestelpa}.  The ability to perform pulse-shape discrimination to distinguish between gamma-ray and thermal neutrons due to different characteristic pulse decay times within the scintillator has also been established~\cite{glodo2011_bestelpa,glodo2012_psd,yang2013_cllb,menge2015_cllb}.

Lanthanum-containing scintillators such as lanthanum-bromide (LaBr$_3$) and lanthanum-chloride (LaCl$_3$) have been shown to exhibit intrinsic radioactive background from naturally occurring $^{138}$La and from alpha decays from actinium contamination~\cite{hartwell2004_lacl,milbrath2004_lacl,kernan2006_la,negm2013_labr}.  Actinium is chemically similar to lanthanum, lying in the same column of the periodic table, making it difficult to separate.  The actinium contamination found in these references appears at electron-equivalent-energies\footnote{The measured energy is commonly given in units of electron-equivalent-energy (\textit{e.g.}, MeVee); a 1~MeV gamma-ray produces 1~MeVee of electron-induced scintillation light, but heavier particles produce less light.} of $\sim$$1.6-3.0$~MeVee, leading to an inseparable large background within a region of interest for gamma spectroscopy.  Yang~\emph{et al.}~\cite{yang2014_labr} found that co-doping cerium-doped LaBr$_3$ with Ca$^{2+}$ or Sr$^{2+}$ significantly increased the electron-equivalent-energy of the alpha background and changed the alpha pulse shapes, enabling pulse-shape discrimination (PSD) to be used to differentiate between gamma- and alpha-induced events.

We have found that CLLB exhibits intrinsic radioactivity that matches the energy spectrum expected from $^{227}$Ac contamination of the lanthanum component.  The estimated activity of the sample we measured was 0.016~Bq/cm$^3$, well below what has been reported in LaBr$_3$ of $0.2-0.8$~Bq/cm$^3$ \cite{yang2014_labr} and LaCl$_3$ of $0.04-0.8$~Bq/cm$^3$ \cite{hartwell2004_lacl}.  We have measured the pulse shapes and energy quenching of the alpha particle interaction and have performed PSD to separate these events from both the gamma and thermal neutron events within CLLB.

\section{Experimental Methods}

We obtained a 1'' right CLLB crystal containing natural lithium on loan from Saint Gobain Crystals that was directly coupled to an R6231-100 photomultiplier tube (PMT).  The PMT was biased to $-1400$~V.  We used an Agilent Acqiris DC282 10-bit waveform digitizer sampling at 2~GS/s to acquire 80,000 waveforms.  Energy spectra and charge integrations were obtained with the LANL-developed Compact Laboratory PSD System (CLPS), which is based on the PSD8C ASIC \cite{engel2009} that was used in two handheld instruments utilizing elpasolites \cite{budden2015_armd, budden2015_caree}.  The CLPS system was designed to be flexible for use with a wide range of elpasolites, allowing the user to input pulse integration window widths, delays, and gains.  This allows a higher data collection efficiency than digitization for studying PSD performance, once ideal integration settings have been found.  We used a $^{137}$Cs gamma-ray source and a moderated $^{241}$Am-B neutron source for both acquisition methods.

In performing pulse-shape discrimination, the figure of merit relates to how well separated two peaks are in a PSD parameter:
\begin{equation}\label{eq:fom}
 \textrm{FOM} = \frac{\mu_A - \mu_B}{\Gamma_A + \Gamma_B}~,
\end{equation}
where $\mu$ is the Gaussian mean and $\Gamma$ the full-width half-maximum (FWHM) of peaks $A$ and $B$ in the PSD parameter.  The optimal PSD parameter depends on the type of elpasolite studied.  For CLLB the average pulse shapes differ the most in the tail region and we chose our PSD parameter to be a ratio of two integration regions we call ``Head'' and ``Full'':
\begin{equation}\label{eq:psd}
 \textrm{PSD ratio} = \frac{\textrm{Head}}{\textrm{Full}}~.
\end{equation}
A FOM greater than $1$ is acceptable, and a FOM for gamma-neutron separation in CLLB as good as 1.9 has previously been reported\footnote{This measurement utilized a different PSD parameter, a ratio of a ``Head'' to ``Tail'' region.}~\cite{menge2015_cllb}.

The digitized waveforms were analyzed to optimize the Head integration region with a fixed 10~$\mu$s Full integration region for the higher statistics CLPS data collection.  Good thermal neutron and gamma separation was observed in the PSD ratio for a Head integration region of $0-445$~ns.  With these integration windows, cuts based on the PSD ratio and the total integrated waveform area (\textit{e.g.} total light output) were applied to select pulses originating from gamma, thermal neutron, and alpha interactions.

\section{Results \& Discussion}

\subsection{Pulse-Shape Discrimination}

\begin{figure}[h!]
 \centering
 \includegraphics[width=0.45\textwidth]{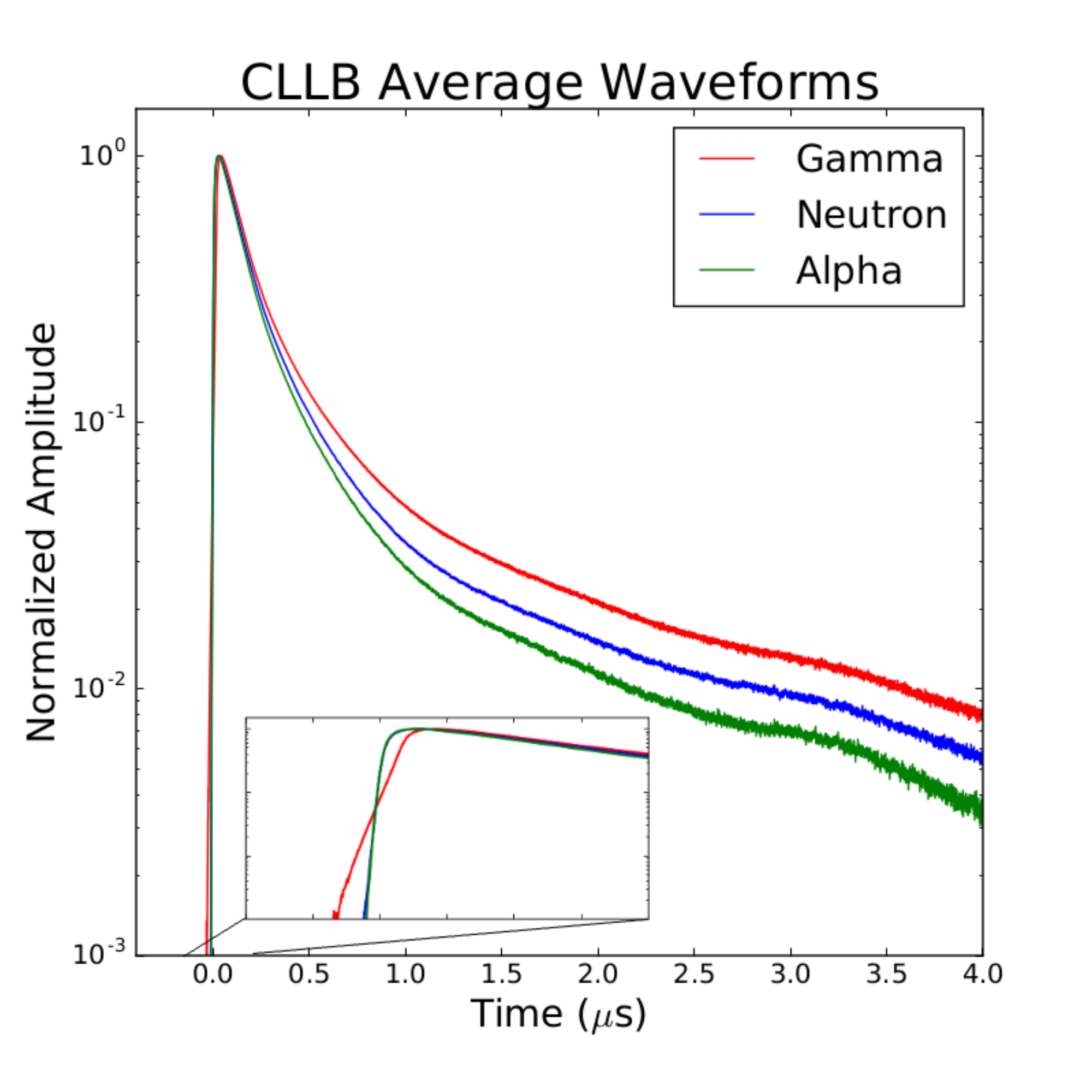}
 \caption{Average waveforms from gamma (red, top curve), thermal neutron (blue, middle curve), and alpha (green, bottom curve) interactions in CLLB.  The inset figure zooms in on the pulse from $-0.1-0.2$~$\mu$s.}
 \label{fig:waves}
\end{figure}

The average pulse shapes we measured for gamma, thermal neutron, and alpha events from CLLB are shown in Fig.~\ref{fig:waves}, normalized to an amplitude of 1.  Thermal neutron capture events result in an alpha particle and a triton, thus the pulse shape for these events reflects a sum of alpha and triton response.  All three pulses exhibit similar rise times: $\sim$24~ns for the gammas and $\sim$15~ns for the thermal neutrons and alphas.  

\begin{figure}[h!]
 \centering
 \includegraphics[width=0.45\textwidth]{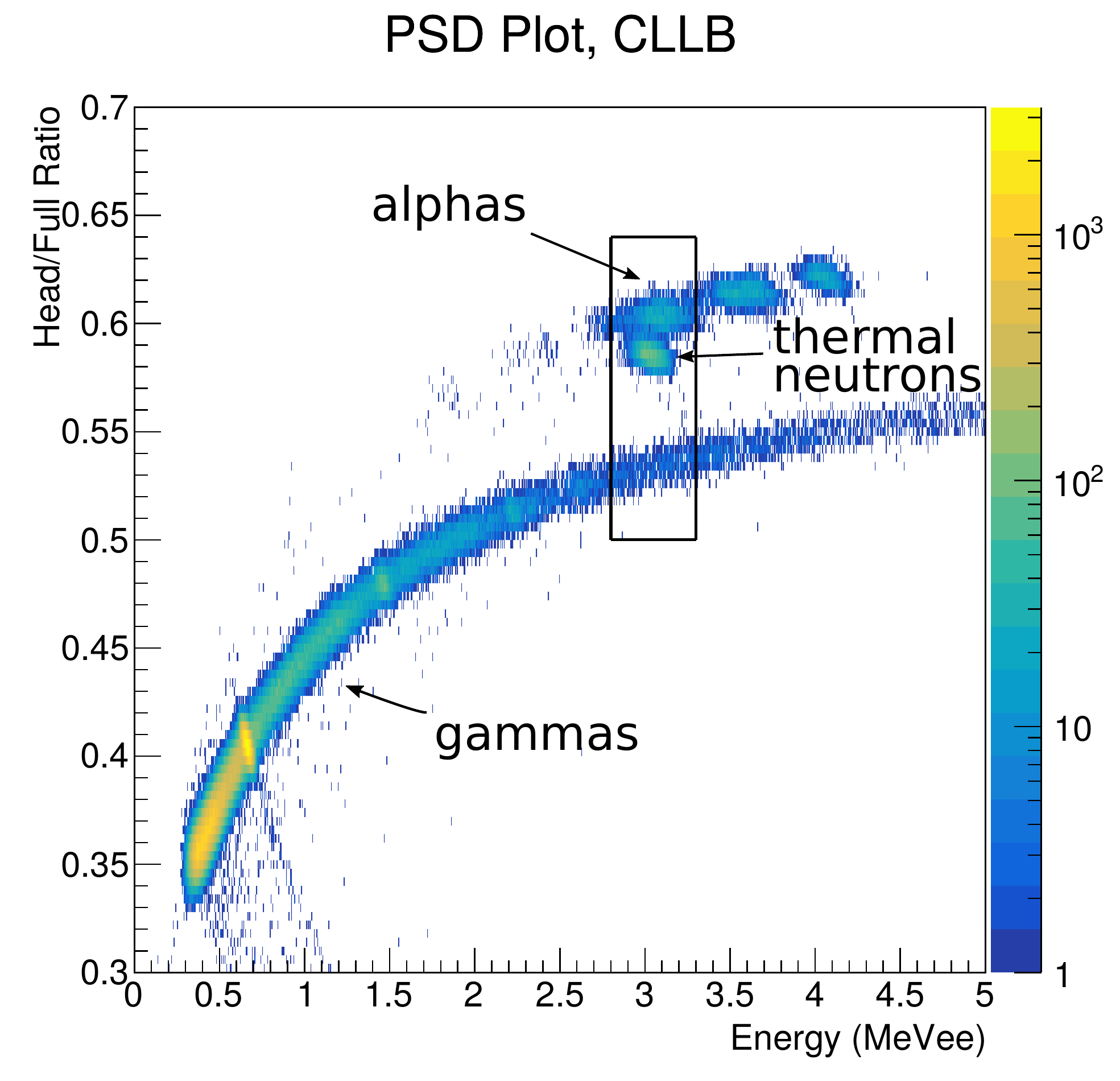}
 \caption{PSD in CLLB, showing a distinct gamma band with low Head/Full ratio and thermal neutron and alpha interactions with higher ratios in the $2.4-4.2$~MeVee region.}
 \label{fig:psd2d}
\end{figure}

Using the integration windows specified and utilizing Eq.~\ref{eq:psd}, the PSD ratio versus electron-equivalent-energy is shown in Fig.~\ref{fig:psd2d}.  The gamma events appear as a band with the lowest PSD Head/Full ratio.  The thermal neutron and alpha events are seen as distinct peaks at higher PSD ratio, and are very well separated from the gamma events.  The alpha events from the intrinsic radioactivity exhibit a larger PSD ratio than the thermal neutron events.  The curvature in the PSD plot is due to a non-zero offset in the integration windows, which allows for better separation between the neutron and alpha events.

\begin{figure}[h!]
 \centering
 \includegraphics[width=0.45\textwidth]{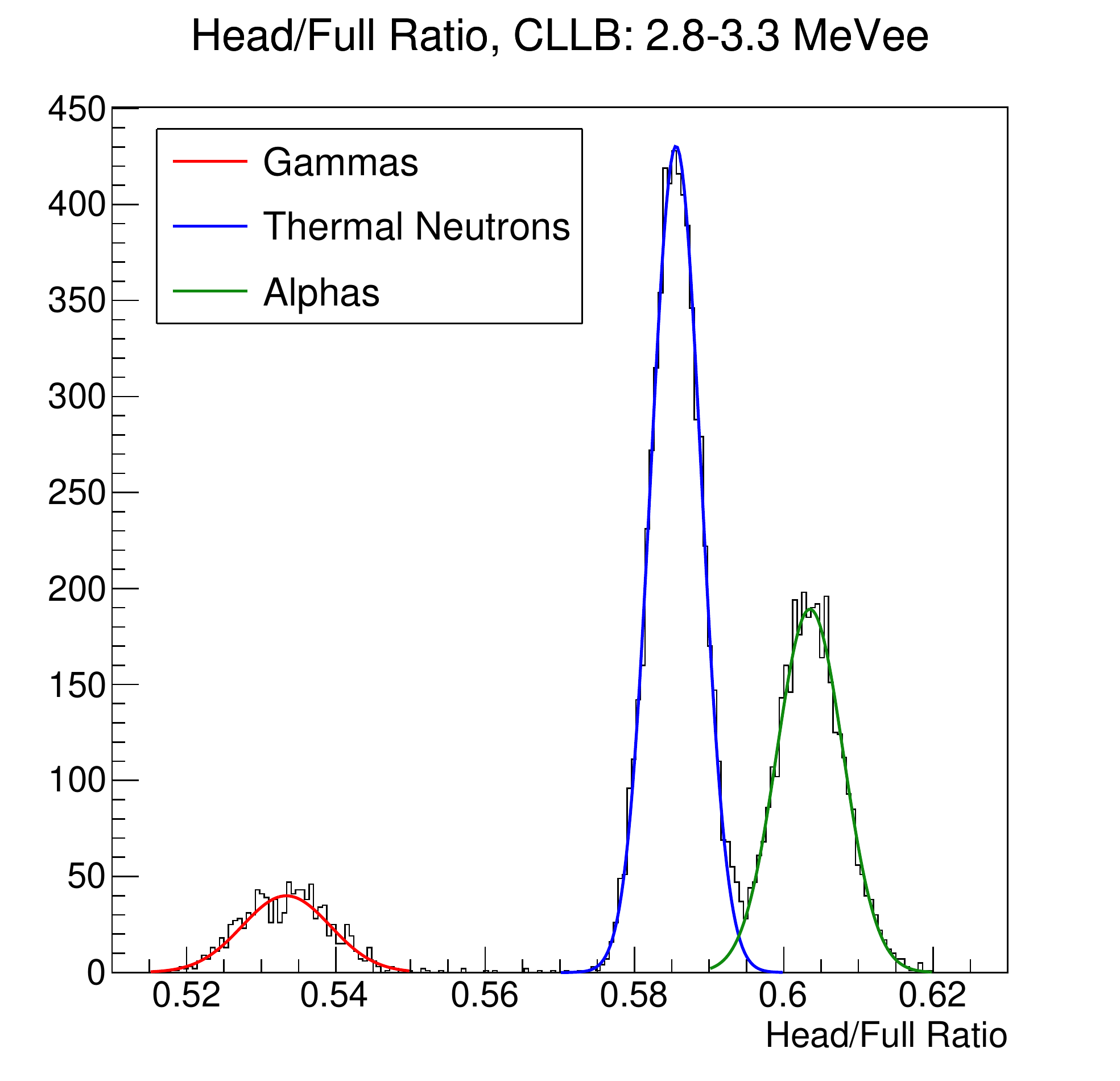}
 \caption{PSD Ratio within $E = 2.8-3.3$ MeVee showing separation for the three particle types, which from left to right are gamma, thermal neutron, and alpha events.}
 \label{fig:psd1d}
\end{figure}

Cutting on the energy region within the box indicated in Fig.~\ref{fig:psd2d}, the PSD ratio is shown in Fig.~\ref{fig:psd1d} illustrating the three distinct peaks which from left to right are associated with gamma, thermal neutron, and alpha events.  Fitting these peaks and utilizing Eq.~\ref{eq:fom}, the FOM between the gamma and thermal neutron events was found to be $2.39 \pm 0.05$ and the FOM between thermal neutron and alpha events was found to be $0.98 \pm 0.03$.

\subsection{CLLB Energy Spectrum}
\begin{figure}[h!]
 \centering
 \includegraphics[width=0.45\textwidth]{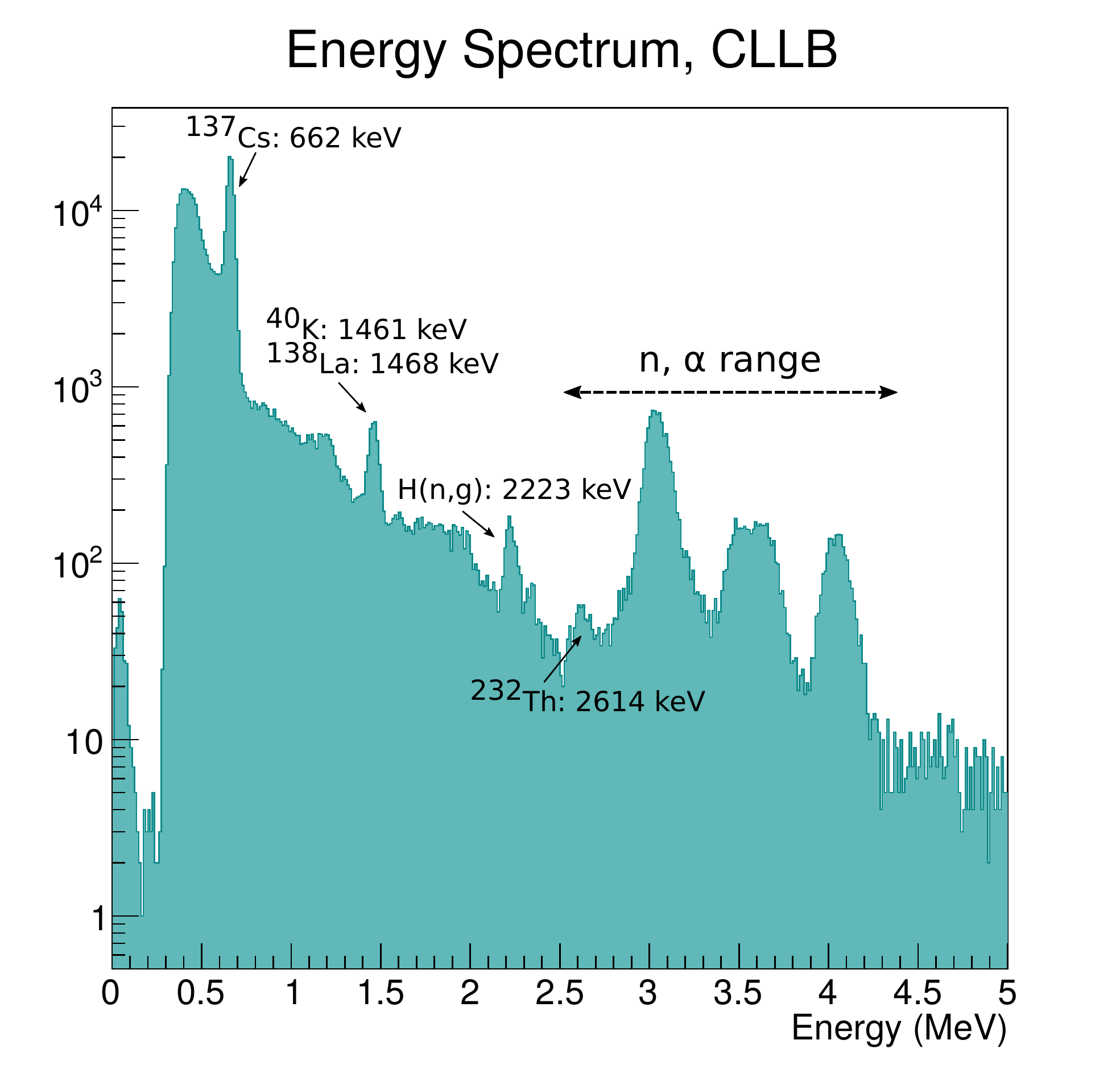}
 \caption{CLLB energy spectrum showing gamma lines from $^{137}$Cs, background, and $^{138}$La radioactivity, thermal neutrons from moderated $^{241}$Am-B, and intrinsic alpha radioactivity.}
 \label{fig:energy}
\end{figure}

The energy spectrum, shown in Fig.~\ref{fig:energy}, was obtained by integrating the pulses over a 10~$\mu$s window.  Two background gamma-ray emission lines, at 1461~keV from $^{40}$K and 2614~keV from $^{208}$Th, can be seen in addition to a 2223~keV gamma emission from neutron capture on hydrogen in our HDPE neutron moderator.  Radioactivity from $^{138}$La, which constitutes $\sim$0.09\% of natural lanthanum, results in a gamma-ray peak near 1468~keV that is the sum of a 1436~keV gamma-ray and coincident electron capture X-rays near 32~keV.  The thermal neutron and alpha events extend over the range of $\sim$$2.4-4.2$~MeVee.  The $^{6}$Li neutron capture events, which have a $Q$-value of 4.8~MeV, appear at an electron-equivalent-energy of 3.03~MeVee, corresponding to a quenching of the scintillation light yield to $63.1\%$ of the light output expected for an electron or gamma of the same energy.

\subsection{Alpha-Particle Quenching}

Actinium-227 is part of the naturally-occurring $^{235}$U decay series and produces alpha particles through alpha decay of five daughters.  Table~\ref{table:chain} shows the alpha particle energies for the different daughter nuclei and corresponding coincident gamma energies for any branching ratio greater than 1\%.  This reaction chain occurs 98.6\% of the time after the $^{227}$Ac $\beta$-decays to $^{227}$Th.  The remaining 1.4\% of reactions are direct $\alpha$-decay of the actinium nucleus, with a decay product energy (alpha particle + coincident $\gamma$-ray) near 4.95~MeV.

\begin{table}[h]
\centering
\caption{Daughter nuclei from $^{227}$Ac that undergo alpha decay, with the associated branching ratios and $\alpha$ and $\gamma$ energies.}\label{table:chain}
\begin{tabular}{|r|c|c|c|}
\hline
Daughter & Branching & $E_\alpha$ & $E_\gamma$\\
 & Ratio (\%) & (MeV) & (MeV) \\
\hline\hline
$^{227}$Th \cite{ndt2001_846}& 2.06 & 5.668 & 0.376 \\
 & 1.50 & 5.693 & 0.351\\
 & 3.63 & 5.701 & 0.343\\
 & 8.30 & 5.709 & 0.334\\
 & 4.89 & 5.713 & 0.330\\
 & 20.4 & 5.757 & 0.286\\
 & 1.27 & 5.808 & 0.235\\
 & 2.42 & 5.867 & 0.175\\
 & 3.00 & 5.960 & 0.079\\
 & 23.5 & 5.978 & 0.061\\
 & 2.90 & 6.009 & 0.030\\
 & 24.2 & 6.038 & -\\
\hline
$^{223}$Ra \cite{ndt2001_802}& 2.22 & 5.434 & 0.445\\
  & 1.00 & 5.502 & 0.376\\
  & 9.00 & 5.540 & 0.338\\
  & 25.2 & 5.607 & 0.270\\
  & 51.6 & 5.716 & 0.159\\
  & 9.00 & 5.747 & 0.127\\
  & 1.00 & 5.871 & -\\
\hline
$^{219}$Rn \cite{ndt2013_2023} & 7.5 & 6.425 & 0.402\\
 & 12.9 & 6.553 & 0.271 \\
 & 79.4 & 6.819 & -\\
\hline
$^{215}$Po \cite{ndt2013_231} & $>$99.9 & 7.386 & -\\
\hline
$^{211}$Bi \cite{ndt2011_707} & 16.2 & 6.278 & 0.351\\
 & 83.5 & 6.623 & -\\
\hline
\end{tabular}
\end{table}

To confirm that the alpha particles measured are in fact from actinium contamination and to determine the amount of light quenching for alpha particles in CLLB we simulated the intrinsic alpha radioactivity and detector response.  The $^{227}$Ac decay chain was simulated by following the decay branching probabilities in a Monte Carlo calculation, which included the associated coincident gamma-rays in determination of the energy spectrum.  The detector response approximated the detector resolution as the FWHM measured at the highest alpha peak (153~keV at 4.05~MeVee) and assumed linearly quenched alphas at these energies, no light from the recoil of the heavy nucleus, and perfect capture of the coincident gammas.  This last assumption we take as a reasonable approximation because the prominent alpha decay branches have no or low-energy coincident gamma rays.  Imperfect collection of the higher energy gammas from lower branching ratio decays will form a Compton background, but will not strongly affect the determination of the peaks.

The quenching relationship we used to determine the measured electron-equivalent-energy of the total scintillation light ($L$) produced in CLLB based on the decay product energies ($E_{\alpha}$, $E_{\gamma}$) takes the form:
\begin{equation}\label{eq:quenching}
 L = E_{\alpha}\times q + L_{0} + E_{\gamma}~,
\end{equation}
where $q$ is the alpha quenching factor and $L_0$ is the value of $L$ at a decay product energy of zero.  A non-zero $L_0$ would be indicative of a nonlinearity in the light quenching at low energies, which has been observed to be the case for organic scintillators \cite{knoll}.  The $^{215}$Po decay peak, which has only one alpha-decay branch and no associated gamma emissions, can be used to relate the quenching factor and offset by $L_0 = L_{Po} - E_{Po}\times q$.  Using this in Eq.~\ref{eq:quenching} reduces the expression to a single-parameter fit of $q$:
\begin{equation}\label{eq:quenching_single}
 L = (E_{\alpha}-E_{Po})\times q + L_{Po} + E_{\gamma}~.
\end{equation}
The energy of the $^{215}$Po decay is $E_{Po} = 7.386$~MeV and it appears in our data peaked at $L_{Po} = 4.05$~MeVee.  

We compared the measured and simulated positions of the alpha peaks and varied the quenching factor $q$ in the simulation.  The simulation and data were fit with four Gaussians based on our resolution: the $^{227}$Th and $^{223}$Ra peaks combined, overlapping $^{211}$Bi and $^{219}$Rn peaks, and the $^{215}$Po peak.  The lowest-energy peak from the $^{227}$Ac alpha decay branch is suggested in our data as a cluster of 30 events around $E = 2.4$~MeVee in Fig.~\ref{fig:psd2d}, but not included in the fit due to low statistics.

\begin{figure}[h!]
 \centering
 \includegraphics[width=0.48\textwidth]{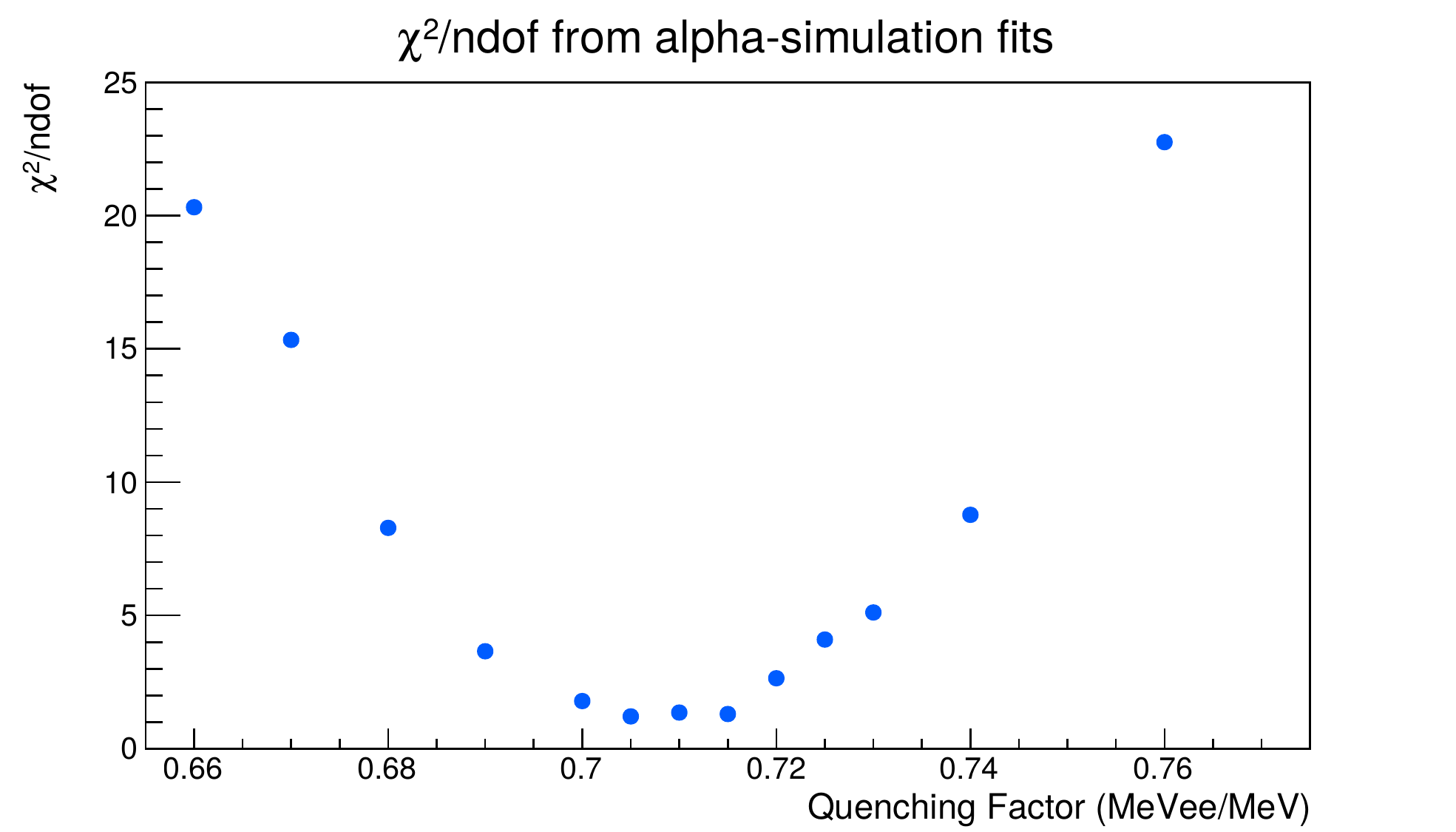}
 \caption{$\chi^2$ from CLLB data compared to our $^{227}$Ac simulation for different quenching factors $q$, showing a minimum at $0.710\pm0.005$.}
 \label{fig:chi2}
\end{figure}

\begin{figure}[h!]
 \centering
 \includegraphics[width=0.45\textwidth]{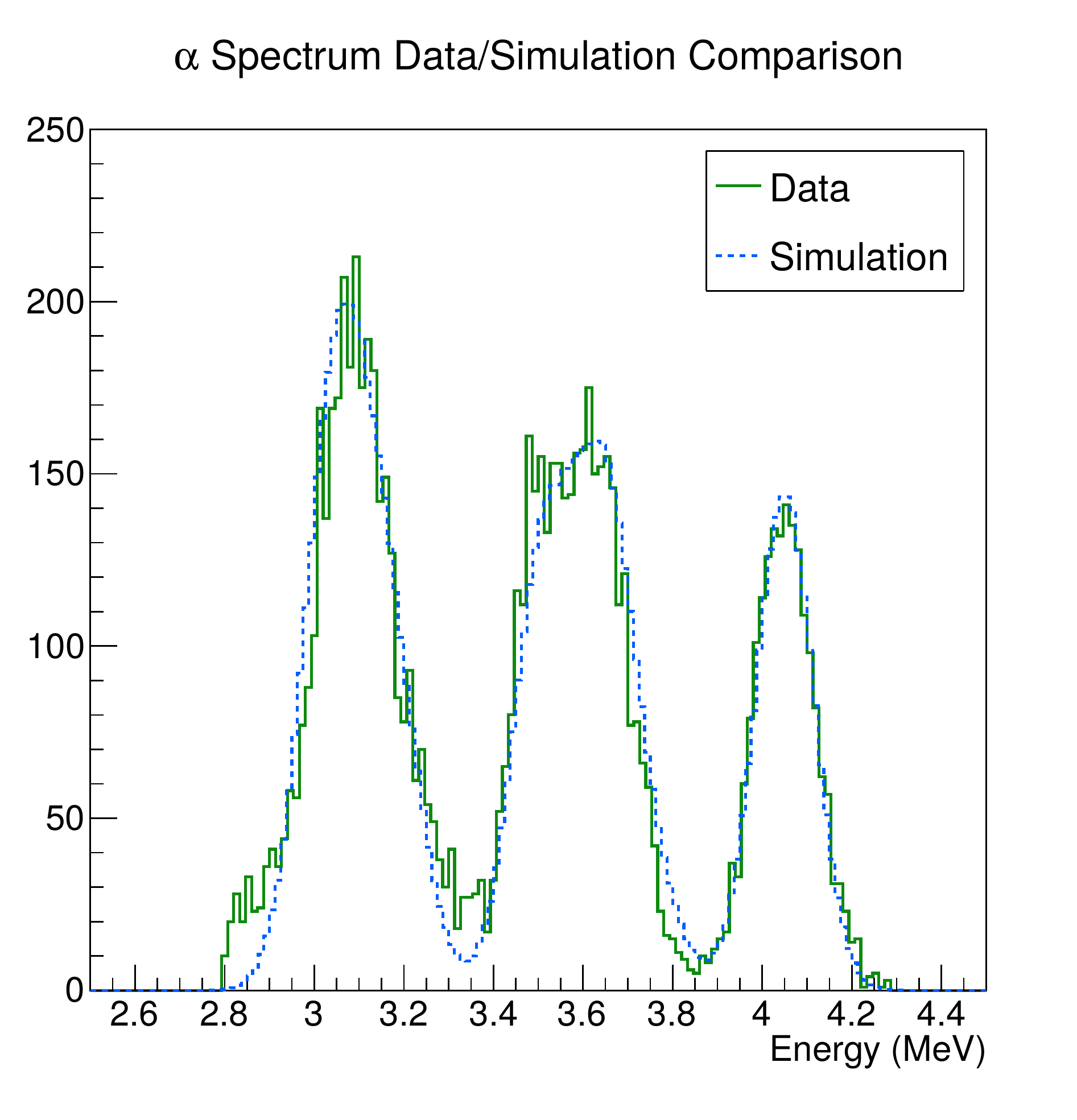}
 \caption{Comparison of CLLB alpha particle spectrum data (green, solid) to simulation (blue, dashed) assuming a quenching factor of 0.71 and using Eq.~\ref{eq:quenching}.}
 \label{fig:simdata} 
\end{figure}

The $\chi^2$ between the fits to measured and simulated data versus the alpha quenching factor is shown in Fig.~\ref{fig:chi2}, indicating a minimum at $q = (0.710\pm0.005)$~MeVee/MeV.  The resulting comparison of our simulation to the data with a quenching factor of $0.71$ is shown in Fig.~\ref{fig:simdata}.  Differences between the simulation and the data in the tail regions of all but the highest-energy peak are indicative of the Compton background we expect to see but did not include in our simulation.

The linear offset associated with this quenching factor is $L_0 = -1.19$~MeVee, thus indicating nonlinearity in the alpha quenching of the light at lower energies.  The agreement between the measured and simulated spectra in Fig.~\ref{fig:simdata} supports the use of a linear quenching relationship in the energy range measured here, between approximately 5.5 and 7.4~MeV. Because the light output of scintillators in response to charged particles is generally nonlinear at low energies and becomes linear at higher energies (\textit{e.g.} \cite{Taylor1951,Bakkum1984,Colonna1992,Slunga2001}), we expect this quenching relationship to continue to be applicable at higher alpha-particle energies to the extent that the narrow energy range  and experimental uncertainties adequately constrain the fit.

We can apply the linear quenching relationship found to the alpha peak from $^{227}$Ac decay which, with only a small 12.9~keV gamma contribution, has an energy of 4.95~MeV.  Our low-statistics measurement of this peak at 2.43~MeVee with a Gaussian $\sigma$ of 0.1 is centered about one standard deviation above the predicted quenched energy of 2.32~MeVee obtained using Eq.~\ref{eq:quenching_single}.  This suggests that the onset of nonlinearity in the quenching likely occurs around 5~MeV.  Equation~\ref{eq:quenching} therefore cannot be used to determine the quenching of the alpha particle from the $^{6}$Li neutron-capture reaction, which has an energy of 2.05~MeV.

\section{Summary}
We have shown that alpha interactions within CLLB exhibit a different pulse shape from the thermal neutron and gamma interactions, thus enabling PSD to discriminate them from the thermal neutrons with a FOM of 0.98 near an energy of 3~MeVee.  Our measured FOM for gamma-neutron discrimination is 2.39, higher than the best previously reported value.  Using PSD to select alpha interactions we measured the alpha-particle energy spectrum from the intrinsic radioactivity in CLLB.  A simulation of the decay chain of $^{227}$Ac folded with our detector response shows that this intrinsic background is consistent with actinium contamination of the lanthanum component.   We determined the linear quenching relationship for the alpha particles above 5~MeV is $L_{\alpha} = E_{\alpha}\times q + L_0$ with $q = 0.71$~MeVee/MeV and $L_0 = -1.19$~MeVee, with a nonlinear quenching response required at lower energies.

\section{Acknowledgements}

The authors would like to thank Anton Zonneveld and Peter Menge at Saint Gobain Crystals for the loan of the CLLB crystal.

These measurements were supported by the Los Alamos National Laboratory Nuclear Nonproliferation and Security Program Office.

\bibliographystyle{elsarticle-num}
\bibliography{biblio}

\end{document}